\documentclass[prb,twocolumn,showpacs, a4paper]{revtex4}
\usepackage{amsmath}
\usepackage{amsfonts}
\usepackage{amssymb}
\usepackage[pdftex]{hyperref}
\usepackage{epsfig}
\usepackage{epstopdf}
\usepackage{graphicx}
\usepackage{enumerate}
\usepackage{multirow}
\usepackage{verbatim}

\begin{document}

\title{$d_{3z^2-r^2}$ orbital in high-$T_c$ cuprates: Excitonic spectrum, metal-insulator phase diagram,  optical conductivity and orbital character of doped holes}
\date{\today}

\author{Xin Wang$^1$, Hung The Dang$^2$, and Andrew J. Millis$^2$}
\affiliation{$^1$Condensed Matter Theory Center, Department of Physics, University of Maryland, College Park, Maryland 20742, USA\\$^2$Department of Physics, Columbia University, 538 West 120$^{th}$ Street, New York, New York 10027, USA}

\begin{abstract}
The single-site dynamical mean-field approximation is used to solve a   model of high-$T_c$ cuprate superconductors which includes both $d_{x^2-y^2}$ and $d_{3z^2-r^2}$ orbitals on the Cu as well as the relevant oxygen states. Both $T$ (with apical oxygen) and $T'$ (without apical oxygen) crystal structures are considered. In both phases, inclusion of the $d_{3z^2-r^2}$ orbital is found to broaden the range of stability of the charge-transfer insulating phase. For equal charge-transfer energies and interaction strengths, the $T'$ phase is found to be less strongly correlated than the $T$ phase. For both structures, $d$-$d$ excitons are found within the charge-transfer gap. However,  for all physically relevant dopings the Fermi surface is found to have  only one sheet and the admixture of $d_{3z^2-r^2}$ into ground state wave function remains negligible ($<5\%$). Inclusion of the extra orbitals is found not to resolve the discrepancy between computed and observed conductivity in the insulating state. \end{abstract}


\maketitle

\section{Introduction}

More than 25 years after their discovery,\cite{Bednorz86} many aspects of the physics of the high-$T_c$ cuprate superconductors remain unclear.\cite{Hanke.rev, Armitage.11} For a long period, researchers attempted to discuss the physics in terms of single-band models, including the $t$-$J$ model and the one-band Hubbard model.\cite{Dagotto94, Lee06, Ogata08} While much of the low-energy physics can be explained by single-band models with appropriately chosen parameters,\cite{Ogata08, Anderson87, Zhang88, Imada.98, Lee98, Anderson04, Yang06} many properties of the cuprates and other transition metal oxides  require consideration of more realistic models.\cite{Dagotto94, Zaanen85, Liebsch.03, Koga.04,Markiewicz.10} The importance of the oxygen bands was stressed early on by Emery and Reiter,\cite{Emery88}  and their ideas were encoded in the ``three-band'' model \cite{Mattheiss87, Emery87, Varma87, Andersen95, Hanke10} which retains the Cu $3d_{x^2-y^2}$ and O $2p_{x,y}$ orbitals on the CuO$_2$ plane. Early qualitative studies of this model\cite{Zaanen85, Kim89, Grilli90, Kotliar91,Dopf92} have been followed by recent quantitative studies\cite{Georges93,Zolfl98,Macridin05,Craco09,demedici09, Wang10a, Wang11} using the dynamical mean-field theory (DMFT)\cite{Georges96,Kotliar06} and sometimes in conjunction with density functional theory calculations.\cite{Weber08,Weber10a,Weber10b} A very recent paper has argued that even the low energy physics may reveal signatures of non-Hubbard or non-$t$-$J$ physics.\cite{Lau11} Although the three-band model helps us in understanding various features of cuprates, it has its limitations. For example, the three-band model has been shown to provide an inadequate description of the optical absorption at frequencies $\omega\gtrsim2$ eV.\cite{demedici09, Wang11} 

A natural question is whether other Cu orbitals, in particular the Cu $3d_{3z^2-r^2}$, play an important role. Higher energy spectroscopies\cite{Nuecker89,Romberg90} have detected these states, which may lead in particular to excitonic states in the spectrum.\cite{Perkins91,Ghiringhelli04,Hancock09,Ament10,Chen10} An early theoretical study,\cite{Feiner92} based on the slave boson approximation, argued that the $d_{3z^2-r^2}$ orbitals are not just admixed into the conduction band, but can give rise to another sheet of the Fermi surface at reasonable doping levels. Variations between material families in the energy and mixing of the $d_{3z^2-r^2}$ orbital were recently argued to affect the value of the second neighbor hopping, thereby explaining the material dependence of $T_c$.\cite{Sakakibara10}  The comparison of theoretical and experimental optical absorption was argued to be at least partially resolved by consideration of the Cu $d_{3z^2-r^2}$ and apical oxygen orbitals.\cite{Weber10b}

These and many other observations motivate this paper, in which we study a six-band model which includes, in addition to the three bands included before, the Cu $3d_{3z^2-r^2}$ orbital and (depending on crystal structure) apical oxygen $2p_z$ orbitals above and below the CuO$_2$ plane. We shall present DMFT calculation of the phase diagram, spectral functions, $d$-$d$ exciton spectrum, optical conductivity and the effect of doping into the $d_{3z^2-r^2}/p_z$ complex. We also study the possible importance of apical oxygen orbitals by comparing the result of $T$-phase (with apical oxygen) and $T'$-phase (without apical oxygen) crystal structures.\cite{Tokura89}
 
The remainder of the paper is organized as follows. In Sec.~\ref{sec:model} we present the model and the method we employed.  Sec.~\ref{sec:phasespec} gives the numerical results of the phase diagram and the spectral functions. Sec.~\ref{sec:ddexc} discusses the $d$-$d$ exciton spectrum. Sec.~\ref{sec:opcond} shows in-plane and $c$-axis optical conductivities, and in Sec.~\ref{sec:FS} we discuss the distribution of hole doping on various orbitals and its consequence on Fermi surfaces.  We conclude in Sec.~\ref{sec:conclusion}.

\section{Model and Method}\label{sec:model}

\begin{figure}[]
    \centering
    \includegraphics[width=0.8\columnwidth, angle=0]{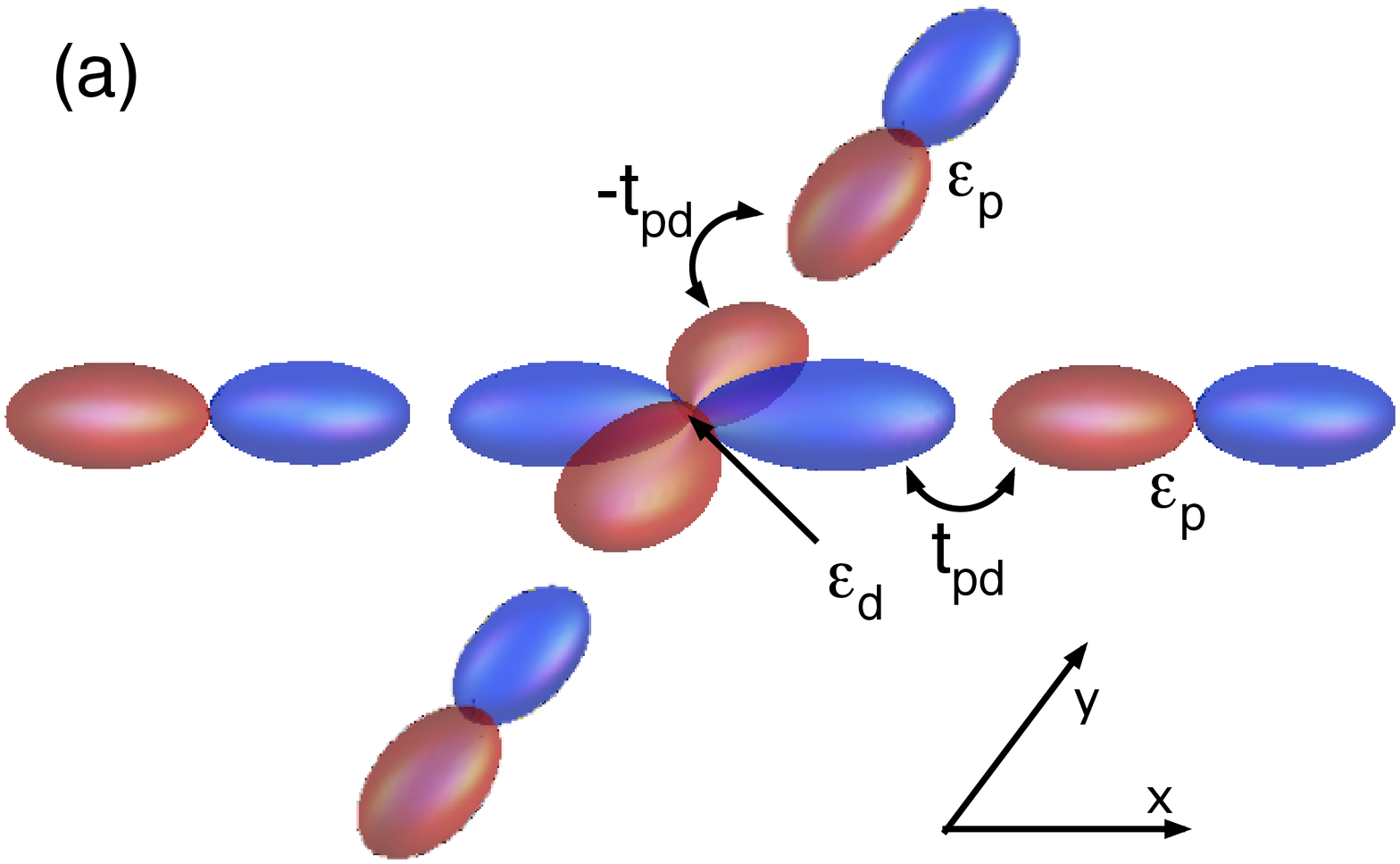}
    \includegraphics[width=0.8\columnwidth, angle=0]{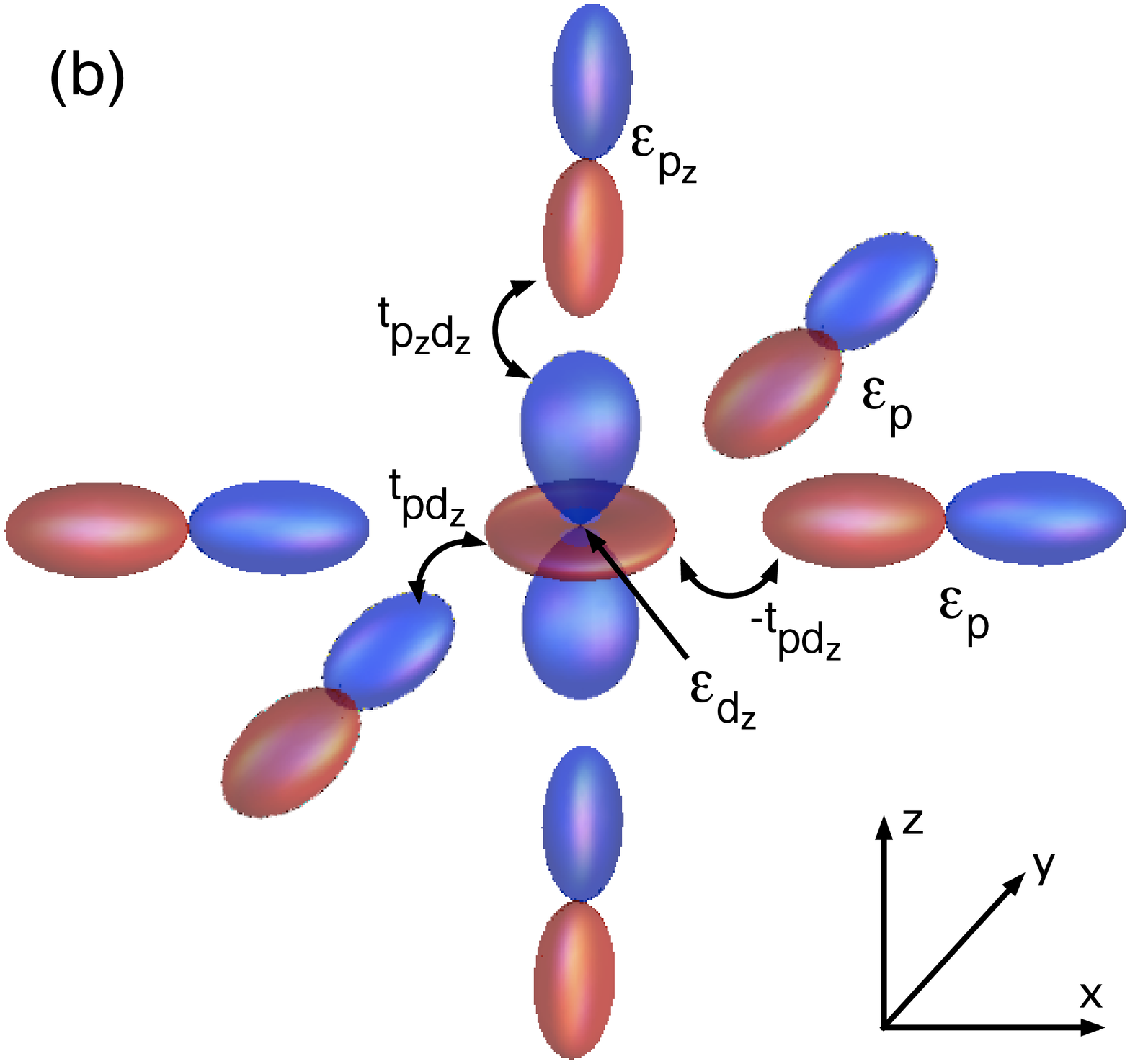}
    \caption{Illustration of orbitals  in the six-band model for the $T$-phase crystal structure. (a)  Cu $d_{x^2-y^2}$ orbital and planar $O_{2p\sigma}$ orbitals which couple to it; (b) Cu $d_{3z^2-r^2}$ orbital and planar and apical orbitals which couple to it.  The different colors (red and blue online) indicate the relative sign of the wave function. In the $T{'}$ phase the apical oxygens are absent. }
    \label{orbitals}
\end{figure}

The three-band model previously considered\cite{Andersen95,demedici09,Wang10a,Wang11} involves the Cu $3d_{x^2-y^2}$ and O $2p_{x,y}$ orbitals pointing to the Cu site in each unit cell. In this work we consider the six-band model, which in addition to the Cu $3d_{x^2-y^2}$ and planar O $2p_{x,y}$ orbitals also includes  the Cu $3d_{3z^2-r^2}$ orbital and two apical oxygen sites (above and below the plane) with one $p_z$ orbital on each site hybridizing with the Cu $3d_{3z^2-r^2}$ orbital. A schematic illustration of orbitals retained in the six-band model is shown in Fig.~\ref{orbitals} with panel (a) showing orbitals relevant to the Cu $3d_{x^2-y^2}$ orbital (which are essentially what included in the three-band model) and panel (b) showing orbitals relevant to the Cu $3d_{3z^2-r^2}$ orbital.  

The resulting model involves six bands, and we restrict attention here to paramagnetic phases, so it is not necessary to write the spin dependence of the band structure explicitly. The band theory part of the Hamiltonian is thus  a $6\times6$ matrix in $\boldsymbol{k}$-space. We neglect the periodicity in the $z$-direction, thus the Hamiltonian only has $k_x$ and $k_y$ dependences. The hopping integrals between Cu and O are also labelled on Fig.~\ref{orbitals}: we use $t_{pd}$ to denote the hopping integral between $p_{x,y}$ and $d_{x^2-y^2}$, $t_{pd_z}$ between $p_{x,y}$ and $d_{3z^2-r^2}$ and $t_{p_zd_z}$ between $p_z$ and $d_{3z^2-r^2}$. Our previous studies of  three-band models shows that the precise value and form of the oxygen-oxygen hopping do not affect the results in any important way.\cite{Wang11} For definiteness, here we obtain estimates for the form and magnitude of  the oxygen-oxygen hopping following  Ref.~\onlinecite{Andersen95}, which argues that the oxygen-oxygen hopping is the result of a virtual process involving hopping on and off the Cu $4s$ orbital. Therefore we derive the six-band model by applying  the L\"owdin downfolding procedure\cite{Loewdin51} to a model involving the six bands considered here plus a Cu $4s$ band (see Appendix for details).  

We use $d_\parallel$ to denote the $d_{x^2-y^2}$ orbital, $d_z$ the   $d_{3z^2-r^2}$  orbital, take the basis $|\psi\rangle=\left({d_{\parallel\boldsymbol{k}}},{d_{z\boldsymbol{k}}}, {p_{x\boldsymbol{k}}}, {p_{y\boldsymbol{k}}},{p_{z\boldsymbol{k}}^{\rm above}}, {p_{z\boldsymbol{k}}^{\rm below}}\right)$  and write the resulting band-theoretic part of the Hamiltonian as
\begin{equation}
{\bf H}_{\rm 6band}=\left(\begin{array}{cc}
{\bf H}_{\rm 6band}^{\rm Cu} & {\bf H}_{\rm 6band}^{\rm hyb} \vspace{0.1cm}\\
\left({\bf H}_{\rm 6band}^{\rm hyb}\right)^\dagger & {\bf
H}_{\rm 6band}^{\rm O}
\end{array}\right),
\end{equation}
where
\begin{equation}
{\bf H}_{\rm 6band}^{\rm Cu}=\left(\begin{array}{cc}
\varepsilon_d & 0\\
0 & \varepsilon_{d_z}
\end{array}\right),
\end{equation}
\begin{widetext}
\begin{equation}
{\bf H}_{\rm 6band}^{\rm O}=\left(\begin{array}{cccc}
\varepsilon_p+2t_{pp}(\cos k_x-1) & -4t_{pp}\sin\frac{k_x}{2}\sin\frac{k_y}{2} & 2it_{pp_z}\sin\frac{k_x}{2} & -2it_{pp_z}\sin\frac{k_x}{2}\\
-4t_{pp}\sin\frac{k_x}{2}\sin\frac{k_y}{2} & \varepsilon_p+2t_{pp}(\cos k_y-1) & 2it_{pp_z}\sin\frac{k_y}{2} & -2it_{pp_z}\sin\frac{k_y}{2}\\
-2it_{pp_z}\sin\frac{k_x}{2} & -2it_{pp_z}\sin\frac{k_y}{2} & \varepsilon_{p_z}-t_{p_zp_z}& t_{p_zp_z}\\
2it_{pp_z}\sin\frac{k_x}{2} & 2it_{pp_z}\sin\frac{k_y}{2} & t_{p_zp_z} &
\varepsilon_{p_z}-t_{p_zp_z}
\end{array}\right),
\end{equation}
\end{widetext}
and the hybridization between Cu and O orbitals:
\begin{equation}
\begin{split}
&{\bf H}_{\rm 6band}^{\rm hyb}\\&=\left(\begin{array}{cccc}
2it_{pd}\sin\frac{k_x}{2} & -2it_{pd}\sin\frac{k_y}{2} & 0 & 0\\
-2it_{pd_z}\sin\frac{k_x}{2} & -2it_{pd_z}\sin\frac{k_y}{2} & t_{p_zd_z}
& -t_{p_zd_z}
\end{array}\right).
\end{split}
\end{equation}
We note that  a linear combination of the two apical oxygen operators decouples from the problem, however for the ease of calculating the $c$-axis conductivity (Sec.~\ref{sec:opcond}) we leave it as it is here, explicitly keeping the two apical oxygen orbitals separately.

We choose $t_{pd}=1.6$eV.\cite{Andersen95} If there is cubic symmetry, $t_{pd_z}=1/\sqrt{3}t_{pd}$ and $t_{p_zd_z}=2/\sqrt{3} t_{pd}$ but in the $T$ phase the  Cu-O bond length is longer along the $z$-axis than $x,y$-axes, resulting in a smaller value of $t_{pd_z}$ and $t_{p_zd_z}$. We follow Ref.~\onlinecite{McMahan90} and use $t_{pd_z}=0.5$eV,  $t_{p_zd_z}=0.8$eV, $t_{pp}=0.6$eV and $t_{pp_z}=0.4$eV. These values are consistent with other estimates found in the literature.\cite{Mila88b,Hybertsen89,Hybertsen90,Korshunov05}  The value of $t_{p_zp_z}$ has not been considered in Ref.~\onlinecite{McMahan90} but since the downfolding procedure implies that $t_{p_zp_z}/t_{pp_z}=t_{pp_z}/t_{pp}$ (see Appendix), we set $t_{p_zp_z}=0.27$eV. We note that the effect of oxygen-oxygen hopping has been studied in detail in Ref.~\onlinecite{Wang11} and it has been shown that the precise values and form of oxygen-oxygen hopping does not change the  physics in any important way. To model the $T'$-phase, in which the apical oxygen states are absent, we set $t_{p_zd_z}=t_{p_zp}=t_{p_zp_z}=0$. 

We next turn to the interaction part of the Hamiltonian. When more than one Cu orbital is important, interactions beyond the Hubbard $U$ must be considered. We adopt the standard Slater-Kanamori form\cite{Slater.36,Kanamori.63} for the interacting part of the Hamiltonian $H_{\rm int}$:
\begin{equation}
\begin{split}
H_{\rm int}=&~U\left(n_{d_\parallel, \uparrow}n_{d_\parallel, \downarrow}+n_{d_z, \uparrow}n_{d_z, \downarrow}\right)\\
&+U'(n_{d_\parallel, \uparrow}n_{d_z, \downarrow}+n_{d_\parallel, \downarrow}n_{d_z, \uparrow})\\
&+(U'-J)(n_{d_\parallel, \uparrow}n_{d_z, \uparrow}+n_{d_\parallel, \downarrow}n_{d_z, \downarrow})\\
&-J\left(d_{\parallel\downarrow}^\dagger d_{z\uparrow}^\dagger d_{z\downarrow}d_{\parallel\uparrow}+d_{z\uparrow}^\dagger d_{z\downarrow}^\dagger d_{\parallel\uparrow}d_{\parallel\downarrow}+h.c.\right)
\end{split}\label{slaterkanamori}
\end{equation} 
Here we have used $d_\parallel^\dagger$ ($d_\parallel$) as the creation (annihilation) operator for the planar $d_{x^2-y^2}$ orbital, and $d_z^\dagger$ ($d_z$) as the creation (annihilation) operator for the $d_{3z^2-r^2}$ orbital. All the interactions are on-site so we have not written the site indices explicitly. We follow the conventional choice of $U'=U-2J$ which comes from symmetry arguments of $d$-orbitals. Note that in keeping with the common practice in modelling cuprates we do not consider interactions on the oxygen sites. At the parameter values we consider the density of holes on the oxygen sites small enough that these interactions are not expected to be important. 

Except in the construction of the phase diagram we will choose the value $U=9$ eV \cite{Mila88,Veenendal94} believed to be representative of cuprates, and set the bare $p$ and $d$ energies equal:  $\varepsilon_d=\varepsilon_{d_z}$ and $\varepsilon_p=\varepsilon_{p_z}$ (where there are $p_z$ orbitals).  We define the bare charge-transfer energy 
\begin{equation}
\Delta=\varepsilon_p-\varepsilon_d.
\label{Deltdef}
\end{equation}

 As will be seen, a large difference in $d$-occupancy and other aspects of the physics arises from difference in in-plane and $c$-axis hopping implied by the crystal structure.    We study $J=0,0.5$ and $1$eV.

We solve the model using the single-site dynamical mean field approximation\cite{Georges96, Kotliar06} with  the continuous-time quantum Monte Carlo impurity solver in its hybridization-expansion (CT-HYB) form.\cite{Werner06b,Werner06a,Gull10}  To obtain real-frequency information we perform analytic continuation of the imaginary-axis self-energies using the method of Ref.~\onlinecite{Wang09}.  The specifics are described in Refs.~\onlinecite{demedici09,Wang11}. The key approximation is assuming that the lattice self-energy is momentum-independent, ${\bf \Sigma}(\omega,\boldsymbol{k})\rightarrow{\bf \Sigma}(\omega)$. The self energy is a matrix in the space of orbitals. Because the interaction is local, which involves only the $d$ electrons,  all entries of ${\bf \Sigma}$ except the $d$-$d$ components vanish.

\section{Phase diagram and spectral functions}\label{sec:phasespec}

\begin{figure}[]
    \centering
    \includegraphics[width=6.8cm, angle=-90]{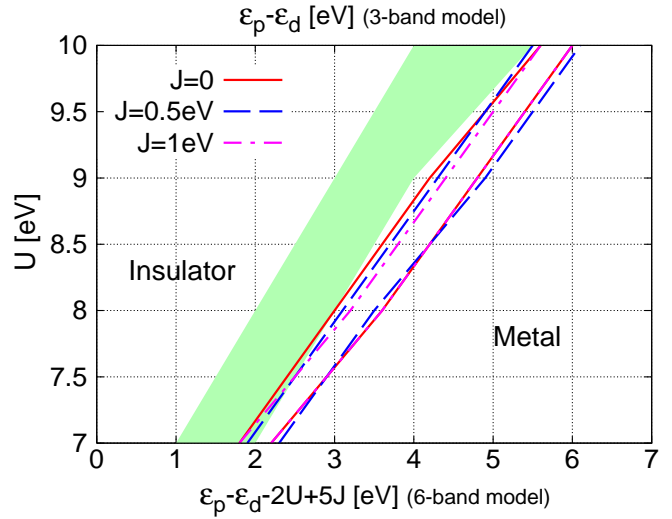}
    \caption{Metal-insulator phase diagram calculated in space of interaction strength $U$ and $p-d$ energy splitting $\varepsilon_p-\varepsilon_d$  at carrier density of one hole per unit cell  for  the six-band model in the $T$-phase at $J$-values indicated (lower $x$-axis, solid, dashed and dash-dotted lines) and compared to the previously published \cite{demedici09} phase diagram for the  three-band model (shaded area, green on-line; upper $x$-axis).  The metal-insulator phase transition is first order, with a region of metastability. The phase diagram shows the limit of stability of the metallic phase, $\Delta_{c2}$ as the left-hand lines (six-band case) or left boundary of shaded region (three-band case) and the limit of stability of the insulating phase,  $\Delta_{c1}$, as the right-hand  lines (six-band case) or right boundary of shaded area (three-band case). }
    \label{phase}
\end{figure}

In this section we present the metal-insulator phase diagram and electron spectral functions for the six-band model for varying choices of Hund interaction $J$ and compare the results to the phase diagram and spectra previously published for the three-band model.\cite{demedici09} To facilitate the comparison  we remove the Hartree energy by shifting the $x$-axis of the six-band model by $-2U+5J$ relative  to the three-band model. The magnitude of the Hartree shift can be understood as follows. In the three-band model the undoped compound is the $d^9$ state with the energy $\varepsilon_d+2\varepsilon_p$; adding one electron leads to the $d^{10}$ state with the energy  $2\varepsilon_d+2\varepsilon_p+U$; the two-hole state nearest in energy is $d^9\underline{L}$ with energy $\varepsilon_d+\varepsilon_p$. Therefore the physical charge-transfer energy is (note that we use electron notation; in some of the literature the charge-transfer energy is defined in hole notation, without the $U$ and with $\varepsilon_d$ and $\varepsilon_p$ reversed)
\begin{equation}
E\left(d^{10}\right)+E\left(d^9\underline{L}\right)-2E\left(d^9\right)=U-(\varepsilon_p-\varepsilon_d)
\end{equation}
However, in   the six-band model there is an additional  Hartree shift arising from the  $3d_{3z^2-r^2}$ orbital. In this case the $d^9$ state has energy $\varepsilon_d+2\varepsilon_{d_z}+2\varepsilon_p+2\varepsilon_{p_z}+3U-5J$ (see, e.g. Table II of Ref.~\onlinecite{Werner06b}); the  $d^{10}$ state has energy $2\varepsilon_d+2\varepsilon_{d_z}+2\varepsilon_p+2\varepsilon_{p_z}+6U-10J$ and the two-hole state nearest in energy is the $d^9\underline{L}$ state whose energy is $\varepsilon_d+2\varepsilon_{d_z}+\varepsilon_p+2\varepsilon_{p_z}+3U-5J$. The physical charge-transfer energy is thus
\begin{align}
 &\ E\left(d^{10}\right)+E\left(d^9\underline{L}\right)-2E\left(d^9\right)\notag\\
= &\ 3U-5J-(\varepsilon_p-\varepsilon_d)\notag\\
= &\ U-(\varepsilon_p-\varepsilon_d-2U+5J)
\end{align}
shifted by $2U-5J$ relative to the three-band model.  The spectra presented below show that six-band models with the same value of $\varepsilon_p-\varepsilon_d-2U+5J$ have the same energy splitting between the  non-bonding oxygen band and the  upper Hubbard band, and that this splitting is also the same as would be found in a three-band model with charge-transfer energy  $\varepsilon_p-\varepsilon_d$.

The solid, dashed and dash-dotted lines in Fig.~\ref{phase} show the phase boundaries calculated from the $T$-phase six-band model for three different values of $J$.  The metal-insulator phase transition is first order \cite{Georges96,demedici09,Wang11} with a coexistence region. $\Delta_{c2}$, the limit of stability of the metallic phase,  is indicated by the left-hand lines in Fig.~\ref{phase}. The limit of stability of the insulating phase is denoted by $\Delta_{c1}$ and is indicated by the right-hand lines. Once the Hartree shift is removed, the Hunds coupling $J$ is seen to have a minor effect on the location of the phase boundary and the width of the crossover regions, although the crossover region is slightly narrower for larger $J$.   

The shaded area (green on-line)  shows  previously published\cite{demedici09} results for the coexistence region of   the three-band model: the left boundary is $\Delta_{c2}$ and the right boundary is $\Delta_{c1}$.  Even after the Hartree shift is removed, the phase boundaries are displaced significantly, and the coexistence regime is wider. Some of the difference in width arises because the three-band model could be studied to lower temperature ($0.025$ eV) than the six-band model, but the difference is larger than the thermal effect.  While a small portion of the difference in location of the phase boundary arises from the difference in Hartree shift arising from small differences in the occupancy of the $d_{x^2-y^2}$ orbitals, the  majority of the change is due to non-Hartree many-body effects.  In essence, in the six-band model the insulating phase remains stable down to weaker values of the effective correlation strength than in the three-band model. We do not have a definitive explanation of this finding at this stage; Further clarification of this issue is important.

\begin{figure}[]
    \centering
    (a)\includegraphics[width=7cm, angle=-90]{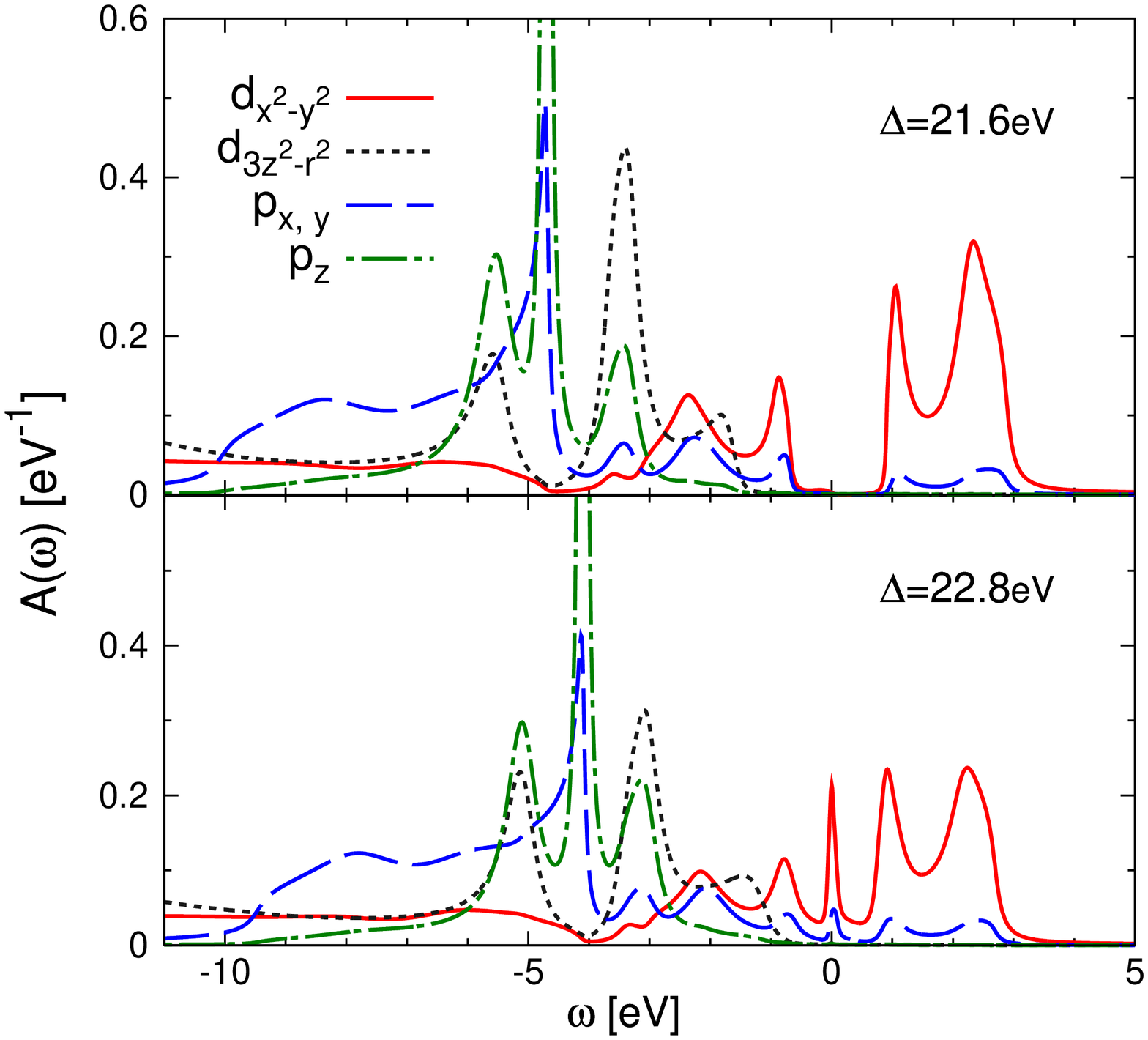}
    (b)\includegraphics[width=7cm, angle=-90]{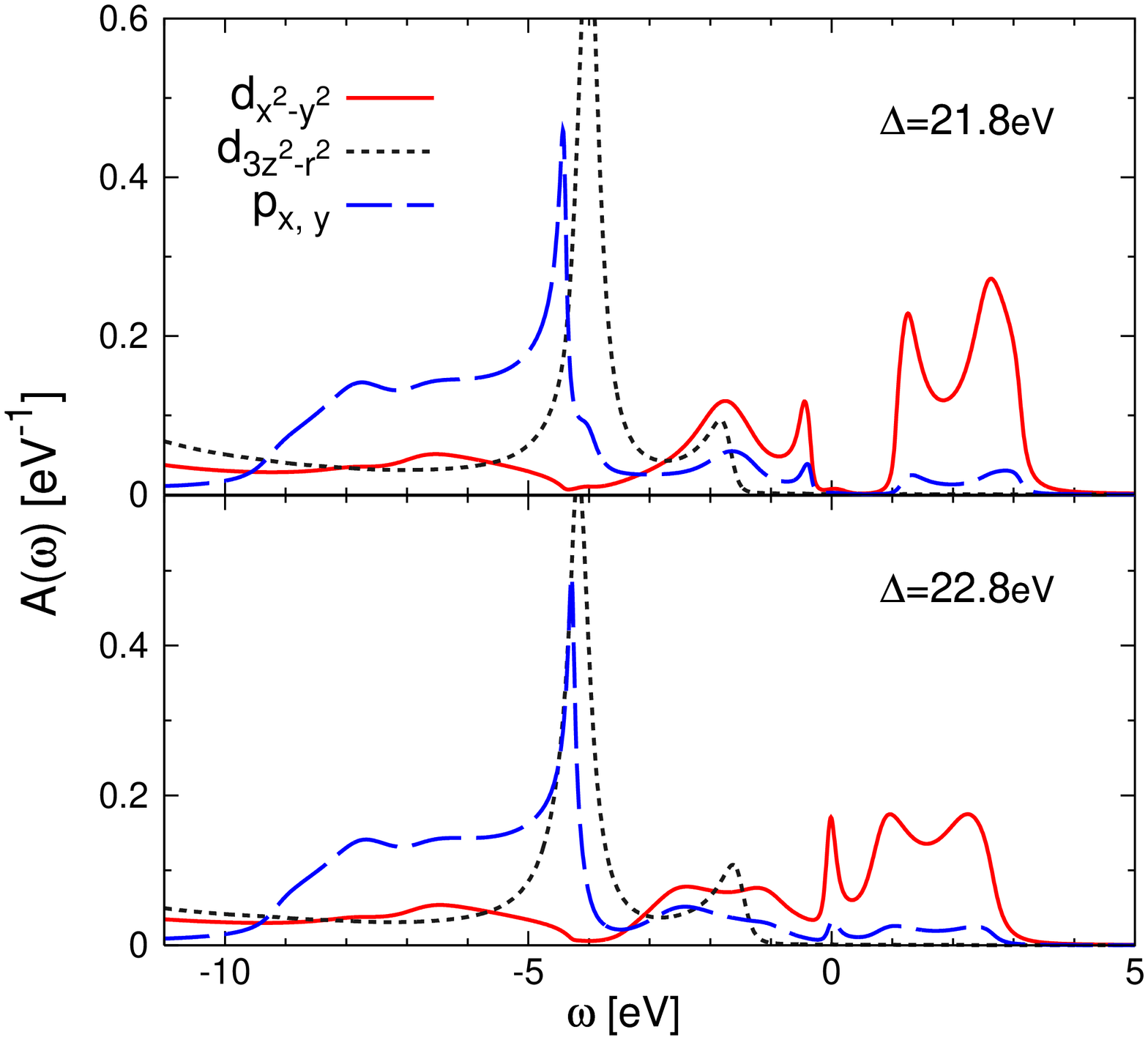}
    \caption{Momentum-integrated spectral functions of the six-band model in the undoped case (one hole per unit cell) for (a) $T$-phase and (b) $T'$-phase. The Fermi energy is at zero. Panel (a): upper part shows the $\Delta=21.6$eV($<\Delta_{c2}$) result and the lower part shows the $\Delta=22.8$eV($=\Delta_{c1}$) result. Panel (b): upper part shows the $\Delta=21.8$eV($<\Delta_{c2}$) result and the lower part shows the $\Delta=22.8$eV($=\Delta_{c1}$) result. Parameters: $U=9$ eV, $J=0$, $T=0.1$ eV. Panel (a) upper part: $\varepsilon_d=\varepsilon_{d_z}=-26.3$ eV, $\varepsilon_p=\varepsilon_{p_z}=-4.7$ eV; lower part $\varepsilon_d=\varepsilon_{d_z}=-26.9$ eV, $\varepsilon_p=\varepsilon_{p_z}=-4.1$ eV. Panel (b) upper part: $\varepsilon_d=\varepsilon_{d_z}=-26.2$ eV, $\varepsilon_p=-4.4$ eV; lower part $\varepsilon_d=\varepsilon_{d_z}=-27.1$ eV, $\varepsilon_p=-4.3$ eV.}
    \label{Aomega}
\end{figure}

This physics is also seen in the spectral functions, presented in Fig.~\ref{Aomega} for the six-band model in the undoped case for the $T$ [panel (a)] and $T'$ [panel (b)] structures at parameters corresponding to the charge-transfer insulator  (upper panels) and paramagnetic metal (lower panels) phases. The results are obtained by maximum-entropy analytic continuation of the self energies, following Ref.~\onlinecite{Wang09}. We note that analytic continuation produces very wide tail down from $-10$ eV which we do not present since it is subject to large uncertainties while being unimportant for our discussion.

The spectra of the $d_{x^2-y^2}$ and the $p_{x,y}$ orbitals are similar to that of the three-band model with comparable parameters.\cite{demedici09,Wang11} The new features are  the $d_{3z^2-r^2}$ and (for the $T$ structure) the apical oxygen $p_z$ orbital. The spectrum of the $p_z$ orbital includes a $\delta$-function centered at $\omega=\varepsilon_{p_z}$ because, as noted above, one linear combination of the $p_z$ orbital decouples.  The two side-bands in the $p_z$ spectrum are the bonding and antibonding portions of the orbital which couple.  Inclusion of oxygen-oxygen hopping between different unit cells in the $z$-direction would broaden the $\delta$-function, however this effect is not important for our considerations.

The hybridization to the $p_z$ orbitals is evident in the spectrum of the $d_{3z^2-r^2}$ orbital: it has mainly a two-peak feature which both at around the same place as the side-bands of the $p_z$ orbitals, although their strengths are quite different. The onset of the $d_{3z^2-r^2}$ spectrum is at a lower energy (around 1 eV) than that of the $d_{x^2-y^2}$ which is due to the fact that the lattice is distorted in the $c$-direction away from the octahedron. The lower part of panel (a) shows the result calculated at $\Delta=\Delta_{c1}$. The ground state is marginally metallic and one can see a narrow quasiparticle peak appears at the Fermi energy. 

Panel (b) of Fig.~\ref{Aomega} shows the result in the $T'$-phase, where the apical oxygen orbitals are absent.  We see similarities in the lineshape of the $d_{x^2-y^2}$ and $p_{x,y}$ orbitals. However the $d_{3z^2-r^2}$ spectrum is quite different: it now has a single peak centered at an energy slightly above $\varepsilon_p$, with its onset similar to panel (a). This is a main change induced by absence of apical oxygen sites. The quasiparticle peak in the lower part of panel (b) is more broad than that of panel (a) indicating that in this case the system is less strongly correlated. Further support for this notion comes from the values of the imaginary part of  Matsubara-axis self energy; here larger magnitudes correspond to larger values of the effective correlation strength. We find, for example, that at a doping of aound $x=0.1$ ${\rm Im} \Sigma$ at the lowest Matsubara frequency is $1.6$eV for the $T$ phase and $1.3$eV for the $T'$ phase.

In constructing the figures we selected values of   $\Delta$ such that $\Delta-\Delta_{c2}$ was the same for the $T$ and $T'$ phase calculations. We can define the charge-transfer energy empirically as the energy difference between the non-bonding oxygen peak and the lowest peak in the upper Hubbard band and the splittings in panels (a) and (b)  are seen to be very similar. Comparison of the upper panels of figures (a) and (b) shows that the gap in the $T'$ phase calcuation is smaller than the gap in the $T$ phase calculation, indicating that for comparable paramters the $T'$ phase is less strongly correlated than the $T$ phase. Comparison of the upper panels of Fig.~\ref{Aomega}(a) and (b) here to Fig.~2(a) of Ref.~\onlinecite{demedici09} shows that a separation of 7 eV leads to metallic behavior in the three-band model but insulating behavior in the six-band model. Examination of data at a distance from the phase boundary in the insulating regime shows that the insulating gap is generically smaller in the six-band case than it is in the three-band case. 

\section{$d$-$d$ exciton spectrum}\label{sec:ddexc}

In this section we discuss the $d$-$d$ exciton spectrum. The corresponding correlation function is defined as:
\begin{equation}
D(\tau)=\left\langle T_\tau[\hat{O}(\tau)\hat{O}^\dagger(0)]\right\rangle\label{dtaudef}
\end{equation}
where the operator $\hat{O}$ is either the singlet exciton operator 
\begin{align}
\frac{1}{\sqrt{2}}\left(d_{\parallel\uparrow}^\dagger d_{z\uparrow}+d_{\parallel\downarrow}^\dagger d_{z\downarrow}\right),
\end{align}
or one of the triplet exciton operators
\begin{align}
d_{\parallel\uparrow}^\dagger d_{z\downarrow},\\
\frac{1}{\sqrt{2}}\left(d_{\parallel\uparrow}^\dagger d_{z\uparrow}-d_{\parallel\downarrow}^\dagger d_{z\downarrow}\right),\\
d_{\parallel\downarrow}^\dagger d_{z\uparrow}.
\end{align}
Here, the $d_\parallel$ and $d_z$ operators have the same meaning as in Eq.~\eqref{slaterkanamori}.

\begin{figure}[]
    \centering
    (a)\includegraphics[width=7cm, angle=-90]{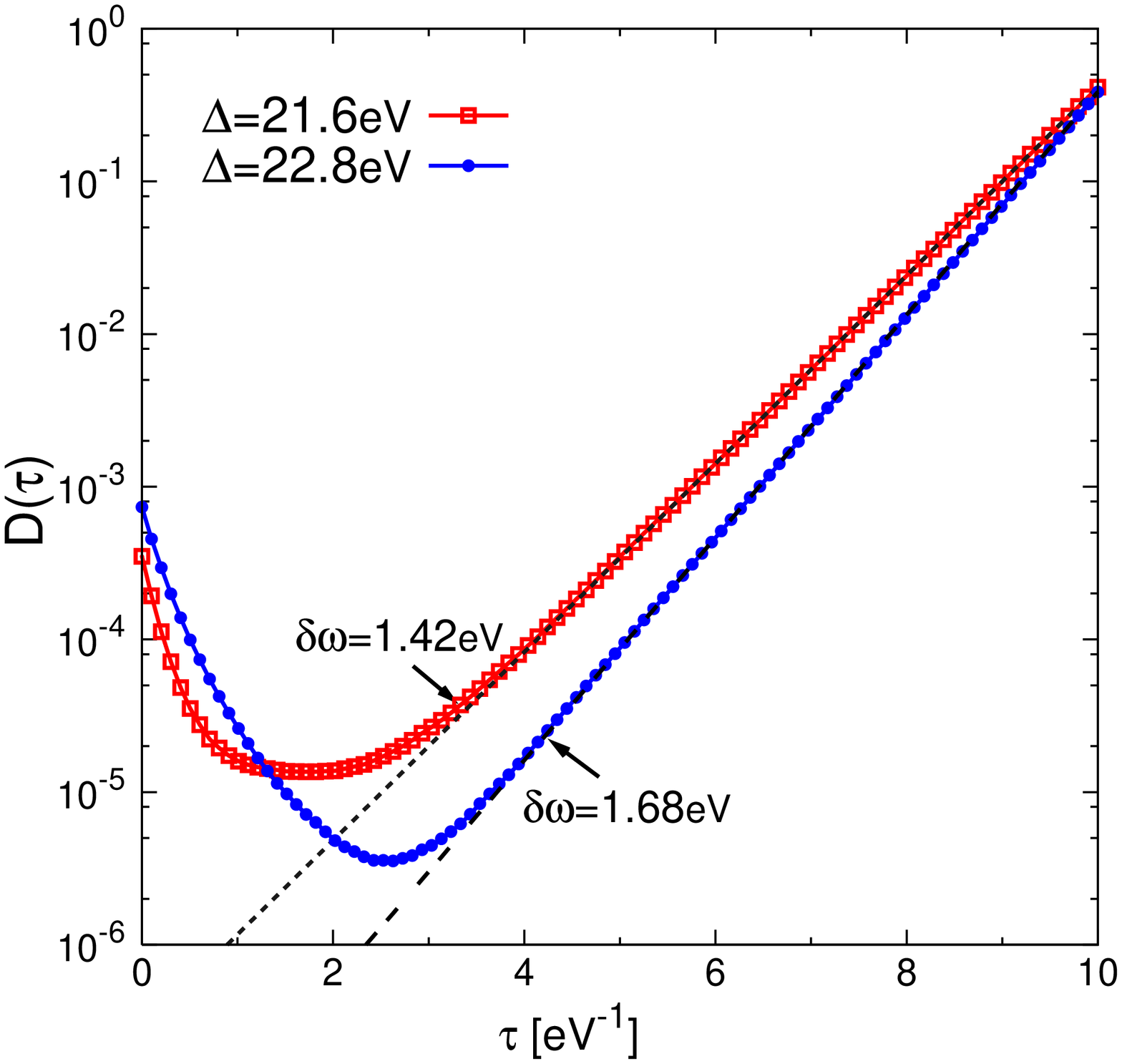}
    (b)\includegraphics[width=7cm, angle=-90]{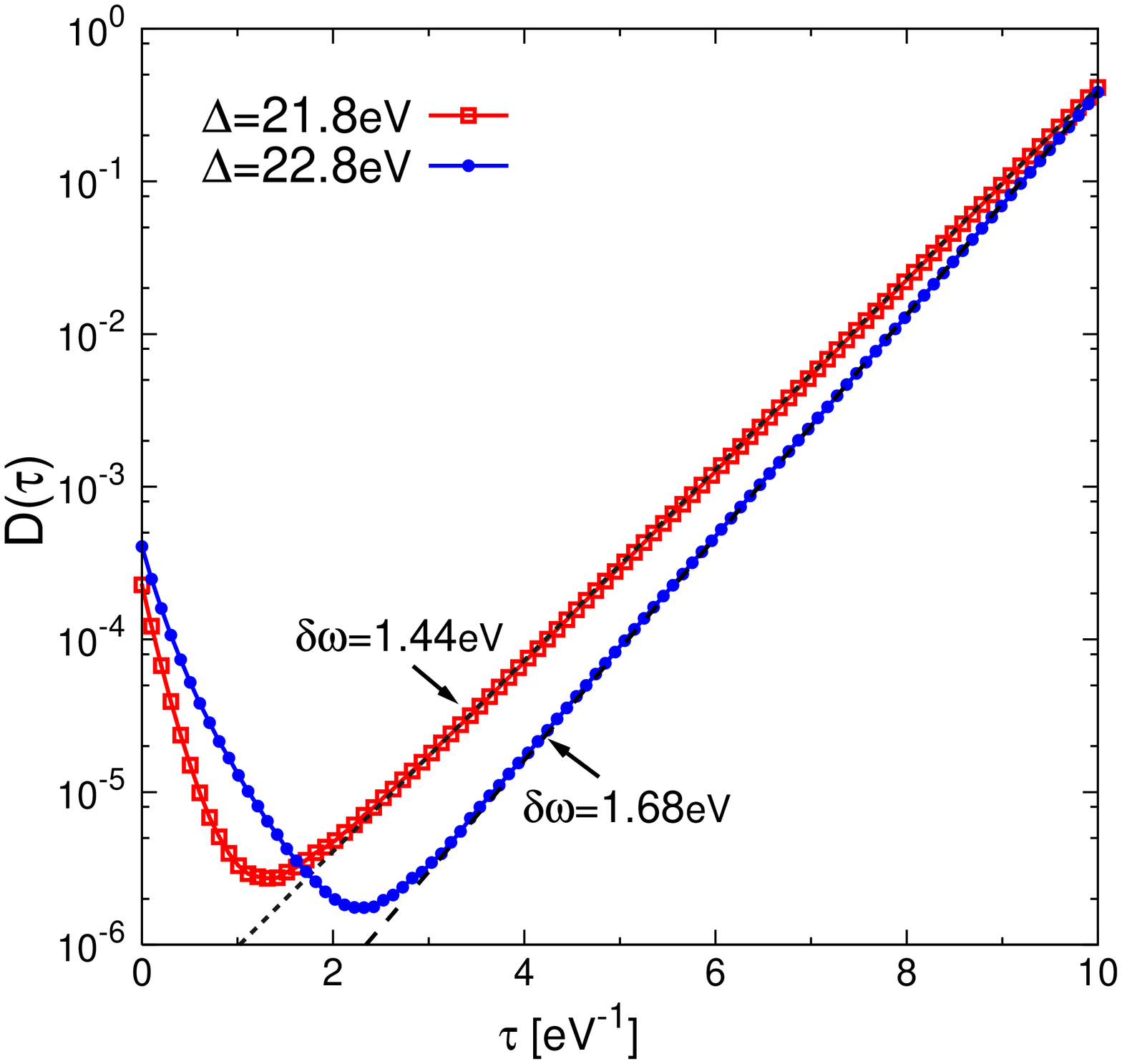}
    \caption{Excitonic correlation functions on imaginary-time axis $D(\tau)$ of the six-band model in the undoped case for (a) $T$-phase and (b) $T'$-phase. Note the semi-log scale. Two different $\Delta$ values with a $\Delta<\Delta_{c2}$ (empty square with lines, red online) and a $\Delta=\Delta_{c1}$ (filled circle with lines, blue online) are shown in each panel. The parameters are the same as in Fig.~\ref{Aomega}. Lines without symbols: a fit to $D(\beta)\exp(-\delta\omega\cdot\tau)$. The fitted $\delta\omega$ is indicated on the figure.}
    \label{ddexctau}
\end{figure}

We have used the CT-HYB procedure to measure $D(\tau)$ [Eq.~\eqref{dtaudef}] in imaginary time. We note that at $J=0$ the singlet and triplet do not split, as expected. However since we are primarily concerned with the one-hole state, even at non-zero $J$ (up to 1 eV) the difference between the singlet and triplet is negligible.  Moreover, we have found (not shown)  that $J$ induces very small effect on the exciton correlation function. We have cross-checked  the lack of $J$-dependence by exactly diagonalizing an isolated CuO$_6$ octahedron. We therefore focus on the $J=0$ results only. Fig.~\ref{ddexctau} shows the results of excitonic correlation functions for the $T$ (upper panel) and $T'$ (lower panel) structures, on the imaginary-time axis  on a semi-log scale for  metallic (blue traces) and insulating (red traces) situations. 

$D(\tau)$ is related to the  real axis spectral function $D(\omega)$ by 
\begin{equation}
D(\tau)=\int d\omega\frac{D(\omega)e^{-\tau\omega}}{1-e^{-\beta\omega}}.\label{dtaudomega}
\end{equation}

In the regime near $\tau=\beta$, $D(\tau)$ drops exponentially, as expected if the real-axis exciton spectrum includes  a $\delta$ function-like peak.  To find the energy of  the peak  we perform a fit of $D(\tau)$ to  $D(\beta)\exp(-\delta\omega\cdot\tau)$ where $\delta\omega$ indicates the binding energy of the peak. The results are shown on Fig.~\ref{ddexctau} with the fitting parameter $\delta\omega$ indicated. We note that although for insulating cases the exciton spectra peaks inside the optical gap, in the metallic cases the exciton has a peak with an even larger gap, meaning that it moves to a slightly higher binding energy. 

The correlation function $D(\tau)$ calculated from CT-HYB is essentially exact: it includes all quantum fluctuations. It is interesting to view this correlation function in diagrammatic terms as a combination of bubble diagram (convolution of interacting Green's function) and vertex correction.  The exciton corresponds to moving an electron from a $d_{3z^2-r^2}$ orbital to a $d_{x^2-y^2}$ orbital. The corresponding bubble diagram is 
\begin{equation}
B_d(i\Omega_n)=-\frac{1}{\beta}\sum_{\omega_n}G_{d_\parallel}(i\omega_n)G_{d_z}(i\omega_n+i\Omega_n).
\end{equation}
or, on the real frequency axis, 
\begin{equation}
B_d(\omega)=\int d\omega'A_{d_\parallel}(\omega')A_{d_z}(\omega+\omega')\left[f(\omega+\omega')-f(\omega')\right]
\end{equation}
where $f(\omega)$ is the Fermi function. $B_d(\omega)$ is the joint $d$-density of states of the $d_{x^2-y^2}$ and $d_{3z^2-r^2}$ orbitals.

It will be useful in our subsequent discussion to   define the total  joint density of states as
\begin{equation}
B_{tot}(\omega)=\int d\omega'A_{tot}(\omega')A_{tot}(\omega+\omega')\left[f(\omega+\omega')-f(\omega')\right]
\end{equation}
where $A_{tot}(\omega)$ is the total spectral function.

\begin{figure}[]
    \centering
    (a)\includegraphics[width=7cm, angle=-90]{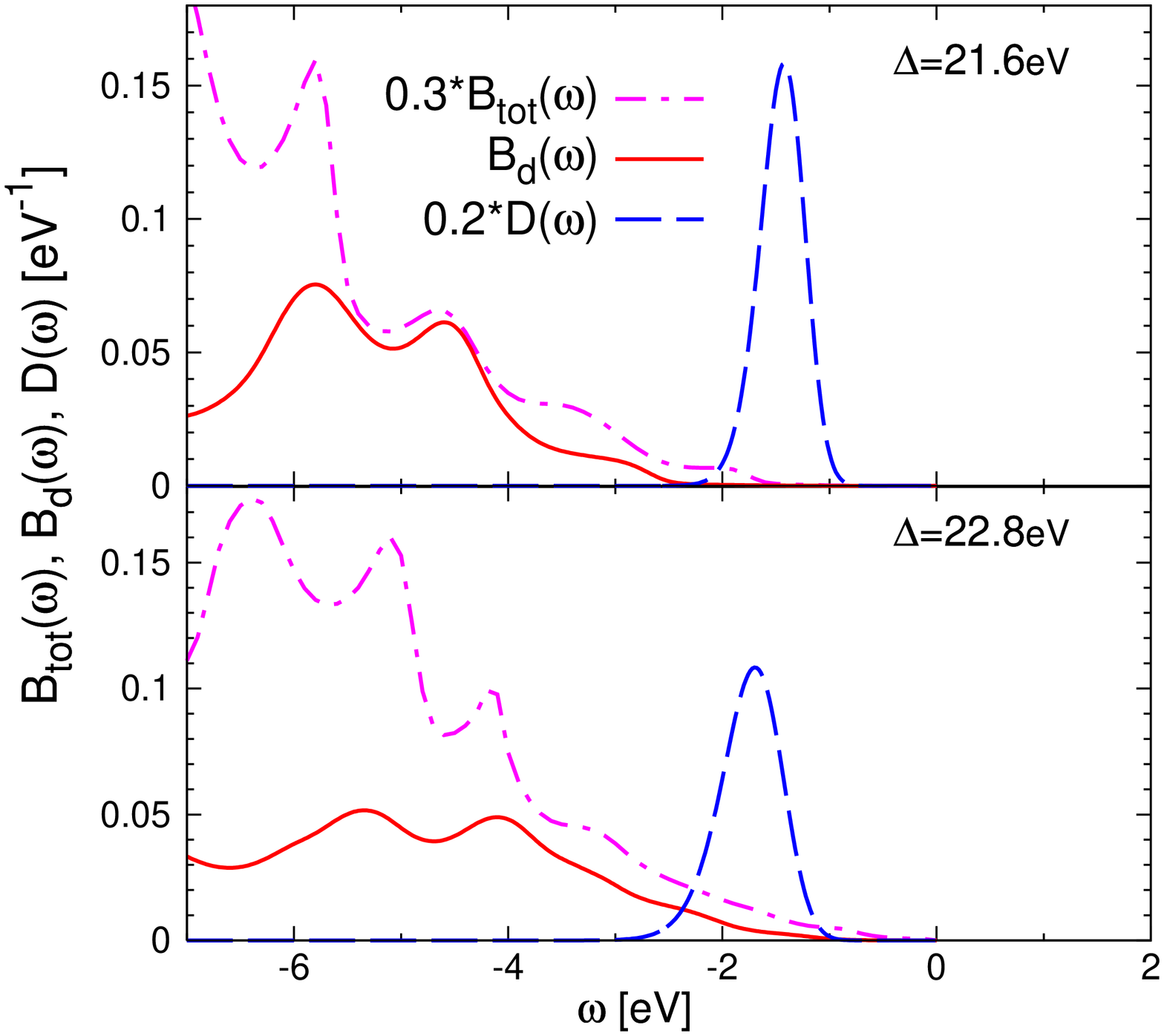}
    (b)\includegraphics[width=7cm, angle=-90]{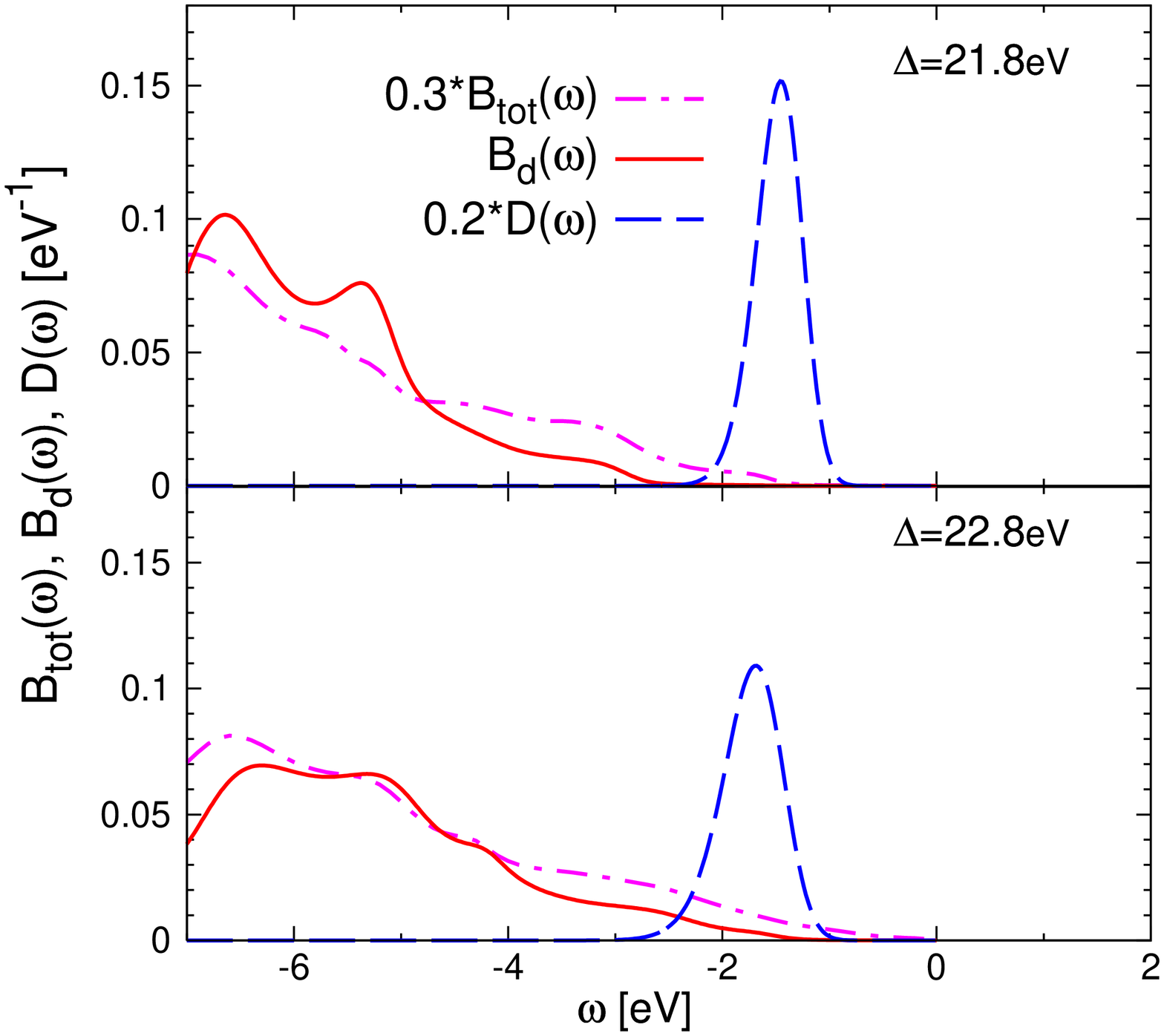}
    \caption{Exciton spectrum (dashed line, blue on-line) calculated for six-band model  at carrier concentration of one hole per unit cell for (a) $T$ structure (b) and $T'$ structure and compared to $d_{3z^2-r^2}/d_{x^2-y^2}$ joint density of states $B_d$ (solid line, red on-line) and total joint density of states $B_{tot}$ (dash-dotted line, magenta on-line). The parameters are indicated on the figure and are the same as in Fig.~\ref{Aomega}. Note that the spectra have been rescaled to facilitate comparison of structures. }
    \label{BDomega}
\end{figure}

In order to compare with these real frequency functions, we have analytically continued the $D(\tau)$ data using the maximum entropy method.\cite{Jarrell.96} Results are presented in Fig.~\ref{BDomega}.  The broadening of $D(\omega)$ is due to the uncertainty of the analytic continuation procedure but the center of the peak is consistent with the exponential fit shown in Fig.~\ref{ddexctau}. It is clear from Fig.~\ref{BDomega} that in the insulating case the exciton spectrum has a peak inside the optical gap, while in the metallic case where the optical gap is closed, the exciton spectrum peak continues to exist as a reasonably well-defined excitation at a slightly higher binding energy.

The exciton energy we find is not consistent with the $0.5$ eV scale proposed in Ref.~\onlinecite{Perkins91} but is reasonably consistent with the discussion in Refs.~\onlinecite{Nuecker89,Romberg90,Ghiringhelli04,Hancock09}.

\section{Optical conductivity}\label{sec:opcond}

In this section we discuss the optical conductivities, in order to determine which features in the optical spectrum may relate to the $d$ orbitals of interest here. Also, previous calculations\cite{demedici09,Wang11} based on the three-band model revealed a strong discrepancy between theory and experiment, with the theoretically calculated conductivity much smaller than the measured one in the region of the charge-transfer gap edge. Ref.~\onlinecite{Weber10b} argued that inclusion of the $d_{3z^2-r^2}$ orbital could resolve this discrepancy. 

The in-plane optical conductivities can be calculated from\cite{Millis.05}
\begin{align}
\sigma(\Omega)&=\frac{2e^2}{\hbar c_0}\int_{-\infty}^\infty \frac{d\omega}{\pi}\int
\frac{d^2\boldsymbol{k}}{(2\pi)^2}\frac{f(\omega)-f(\omega+\Omega)}{\Omega} \nonumber
\\
&\times{\mathrm{Tr}}\left[{\bf j}(\boldsymbol{k}){\bf A}(\omega+\Omega,\boldsymbol{k}){\bf
j}(\boldsymbol{k}){\bf A}(\omega,\boldsymbol{k})\right], \label{sigmamatrix}
\end{align}
where $c_0$ is the $c$-axis lattice parameter, $f(\omega)$ is the Fermi function, the $\boldsymbol{k}$-integral is over the full Brillouin zone with $k$ scaled to $\pi$ divided by the in-plane lattice parameter $a$, and 
$
{\bf A}(\omega,\boldsymbol{k})=
\left[{\bf G}(\omega,\boldsymbol{k})-{\bf G}^\dagger(\omega,\boldsymbol{k})\right]/(2i)$ is the electron spectral function, a matrix in orbital space. We have chosen our  Fourier transform so that the in-plane  current operator is simply a $\boldsymbol{k}$-derivative of the Hamiltonian matrix, ${\bf j}=\partial{\bf H}/\partial k_x$;\cite{Wang11} the extra terms discussed by Ref. \onlinecite{Tomczak09} are not needed.

The $c$-axis optical conductivity can be calculated in a similar manner using the current operator
\begin{widetext}
\begin{equation}
{\bf j}_{\rm c}=
\left(\frac{c_0}{a}\right)^2\left(\begin{array}{cccccc}
0 & 0 & 0 & 0 & 0 & 0\\
0 & 0 & 0 & 0 & -\frac{i}{2}t_{p_zd_z} & -\frac{i}{2}t_{p_zd_z}\\
0 & 0 & 0 & 0 & t_{pp_z}\sin\frac{k_x}{2} & t_{pp_z}\sin\frac{k_x}{2}\\
0 & 0 & 0 & 0 & t_{pp_z}\sin\frac{k_y}{2} & t_{pp_z}\sin\frac{k_y}{2}\\
0 & \frac{i}{2}t_{p_zd_z} & t_{pp_z}\sin\frac{k_x}{2} & t_{pp_z}\sin\frac{k_y}{2} & 0 & it_{p_zp_z}\\
0 & \frac{i}{2}t_{p_zd_z} & t_{pp_z}\sin\frac{k_x}{2} & t_{pp_z}\sin\frac{k_y}{2} & -it_{p_zp_z}& 0
\end{array}\right),
\label{Jcaxis}
\end{equation}
\end{widetext}
and in  La$_2$CuO$_4$ $c_0/a\sim 1.3$.

\begin{figure}[]
    \centering
    \includegraphics[width=7cm, angle=-90]{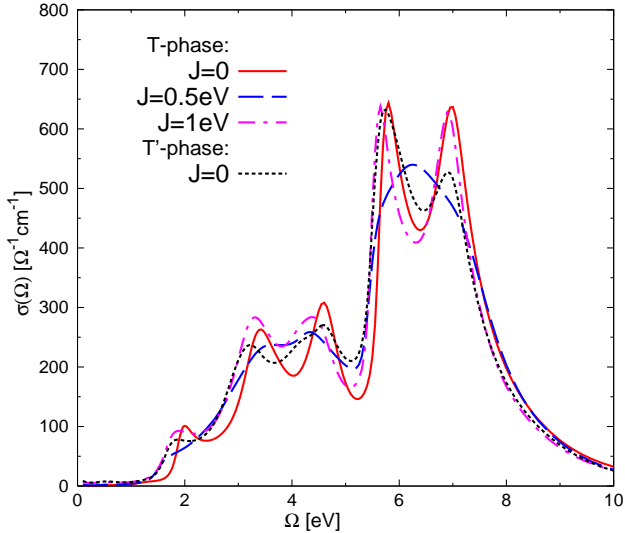}
    \caption{In-plane optical conductivities calculated for $T$- and $T'$-phase in the undoped case at $\Delta$ values with similar distances below the insulating boundary $\Delta_{c2}$. $U=9$ eV and $T=0.1$ eV. Parameters: $T$-phase: $J=0$ (red solid line), $\Delta=21.6$ eV, $\varepsilon_d=\varepsilon_{d_z}=-26.3$ eV, $\varepsilon_p=\varepsilon_{p_z}=-4.7$ eV.
    $J=0.5$ eV (blue dashed line), $\Delta=19.2$ eV, $\varepsilon_d=\varepsilon_{d_z}=-23.8$ eV, $\varepsilon_p=\varepsilon_{p_z}=-4.6$ eV.
     $J=1$ eV (magenta dash-dotted line), $\Delta=16.8$ eV, $\varepsilon_d=\varepsilon_{d_z}=-21.3$ eV, $\varepsilon_p=\varepsilon_{p_z}=-4.5$ eV.
$T'$-phase (black dotted line): $J=0$, $\Delta=21.8$ eV, $\varepsilon_d=\varepsilon_{d_z}=-26.2$ eV, $\varepsilon_p=-4.4$ eV.
    }
    \label{inplanecond}
\end{figure}

Fig.~\ref{inplanecond} shows the calculated in-plane optical conductivity for the $T$ and $T'$ phases. In the two calculations  the $\Delta$ values are chosen to be at approximately the same distance from the insulating phase boundary $\Delta_{c2}$ so the  gap sizes are quite similar. The two systems give very similar in-plane conductivities. Further, the results are very similar to those obtained from the three-band model\cite{demedici09,Wang11}: an onset of absorption above around 1.8 eV and a strong absorption at energy between 6 eV and 8 eV. The rise in the 3 eV-5 eV range is due to the effect of the $d_{3z^2-r^2}$ orbital. The introduction of additional orbitals does not increase the conductivity magnitude in the  frequency range immediately above the gap ($\Omega\sim 2-3$eV) significantly: the disagreement with experiment previously noted in the three-band model \cite{demedici09,Wang11} remains. These results do not agree with results presented in Ref.~\onlinecite{Weber10b}.

\begin{figure}[]
    \centering
    \includegraphics[width=7cm, angle=-90]{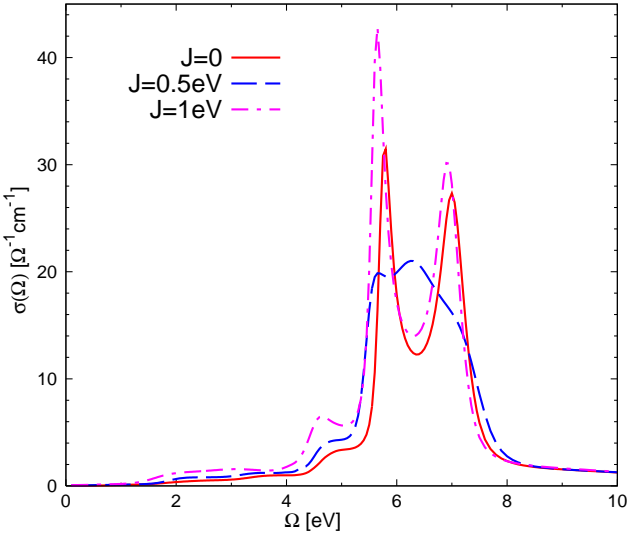}
    \caption{$c$-axis optical conductivities calculated for the $T$-phase in the undoped case at $\Delta$ values with similar distances below the insulating boundary $\Delta_{c2}$. Parameters are the same as in Fig.~\ref{inplanecond}.
    }
    \label{caxiscond}
\end{figure}

Fig.~\ref{caxiscond} shows the results of the $c$-axis optical conductivity calculated in the $T$-phase. The overall magnitude is at least an order of magnitude smaller than that of the in-plane conductivity, which is a consequence of a much smaller hybridization strength and smaller number of relevant orbitals in the $c$-direction. In the 2 eV to 4 eV range the conductivity is non-zero but very small. At 4 eV the conductivity start to rise, signalling the onset of transitions involving  the $d_{3z^2-r^2}$ orbitals. Between 6 eV and 8 eV there are two strong peaks which we consider to be the transition between the upper Hubbard band (which has a double peak structure) and the non-bonding $p_z$ band.

\section{Orbital character of doped holes and shape of Fermi surfaces}\label{sec:FS}

In this section we consider the variation with doping of the orbital character of the low-lying states. This section is motivated by the possibility that above a critical doping the $d_{3z^2-r^2}$ band begins to become occupied. 

\begin{figure}[]
    \centering
    \includegraphics[width=7cm, angle=-90]{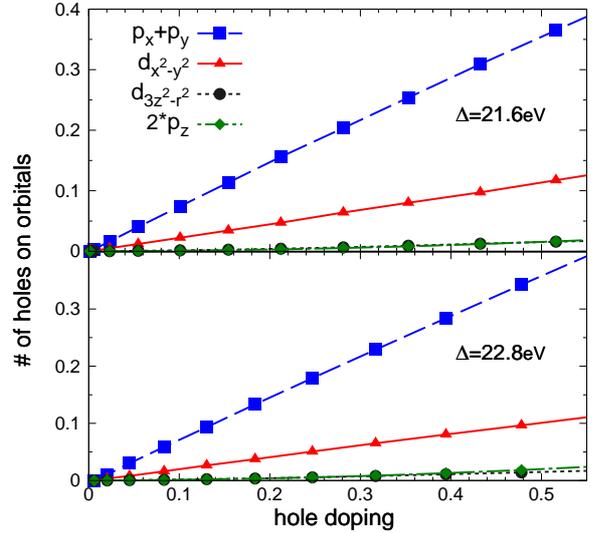}
    \caption{Number of holes on each orbital as a function of the total hole doping per unit cell, calculated for the six-band model in $T$-phase. Upper panel: $\Delta=21.6$ eV ($<\Delta_{c2}$). Lower panel: $\Delta=22.8$ eV ($=\Delta_{c1}$). Note that the number of holes on $p_x$ and $p_y$ orbital of the in-plane oxygen sites are combined as $p_x+p_y$ (shown as blue squares with long dashed lines), and the number of holes on the $p_z$ orbital of the above- and below-plane apical oxygen sites are similarly combined as $2*p_z$ (shown as green diamonds with short dashed lines). Therefore at a given doping value the sum over the value at the four curves gives the correct total hole doping per unit cell.}
    \label{numholes}
\end{figure}

Fig.~\ref{numholes} shows the doping dependence of the number of holes on each orbital per unit cell. From the spectral functions shown in Fig.~\ref{Aomega} one would expect that the number of holes on the $d_{3z^2-r^2}$ orbital will dramatically increase when the chemical potential is reduced below a certain point. The theoretical arguments of Ref.~\onlinecite{Feiner92} also suggest that this will occur. However the actual DMFT calculation is inconsistent with the rigid band picture. We see that while the total number of holes in the d-band increases linearly with doping, the hole density  on the $d_{3z^2-r^2}/p_z$ complex remains very  small even at very high doping levels. Thus the  spectra deform as  the chemical potential is reduced, in such a way that the  $d_{3z^2-r^2}$ orbital remains far below the Fermi level.

\begin{figure}[]
    \centering
    \includegraphics[width=8cm, angle=0]{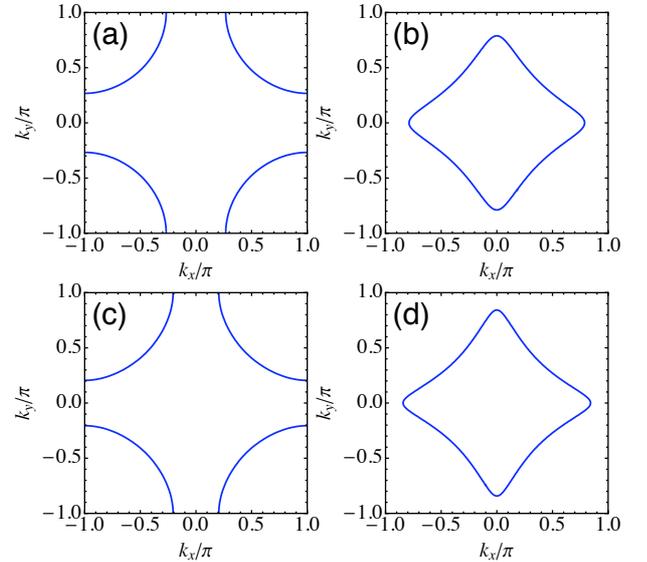}
    \caption{Fermi Surfaces of the six-band model in $T$-phase. Panel (a): $\Delta=21.6$ eV ($<\Delta_{c2}$), hole doping $x=0.10$. Panel (b): $\Delta=21.6$ eV, $x=0.35$. Panel (c): $\Delta=22.8$ eV ($=\Delta_{c1}$), $x=0.08$. Panel (d): $\Delta=22.8$ eV, $x=0.32$. Parameters: $U=9$ eV, $T=0.1$ eV. 
    Panel (a): $\varepsilon_d=\varepsilon_{d_z}=-25.3$ eV, $\varepsilon_p=\varepsilon_{p_z}=-3.7$ eV.
     Panel (b): $\varepsilon_d=\varepsilon_{d_z}=-24.5$ eV, $\varepsilon_p=\varepsilon_{p_z}=-2.9$ eV.
      Panel (c): $\varepsilon_d=\varepsilon_{d_z}=-26.3$ eV, $\varepsilon_p=\varepsilon_{p_z}=-3.5$ eV.
       Panel (d): $\varepsilon_d=\varepsilon_{d_z}=-25.5$ eV, $\varepsilon_p=\varepsilon_{p_z}=-2.7$ eV.}
    \label{fermisurface}
\end{figure}

To gain further insight into the doping dependence we plot the Fermi surfaces of the six-band model in $T$-phase in Fig.~\ref{fermisurface}. Panels (a) and (b) shows results obtained for parameters such that at half filling the model is in its paramagnetic insulating phase while panels (c) and (d) show results obtained for parameters such that at half filling the model is in the paramagnetic metallic phase.  The hole doping values of panels (a) and (c) are selected around 0.1 and panels (b) and (d) around 0.35. We see that the Fermi surface includes only one sheet in all cases, consistent with the discussion above that the crossing of the Fermi energy into the $d_{3z^2-r^2}$ band is avoided. For the smaller doping value the Fermi surface is approximately a circle enclosing $(\pi,\pi)$ and for the larger doping the Fermi surface is star shaped enclosing the $(0,0)$ point. Thus, in disagreement with early slave boson studies,\cite{Feiner92} we find that in the six-band model there is no physically relevant doping at which holes occupy $d_{3z^2-r^2}/p_z$ orbitals as separate bands, and the Fermi surface remains one-sheeted. However, we do note that the van Hove singularity happens at around hole doping value $x\approx0.28$, an intermediate value between what shown in panels (a), (c) and (b), (d) in Fig.~\ref{fermisurface}.

\begin{figure}[]
    \centering
    \includegraphics[width=7cm, angle=-90]{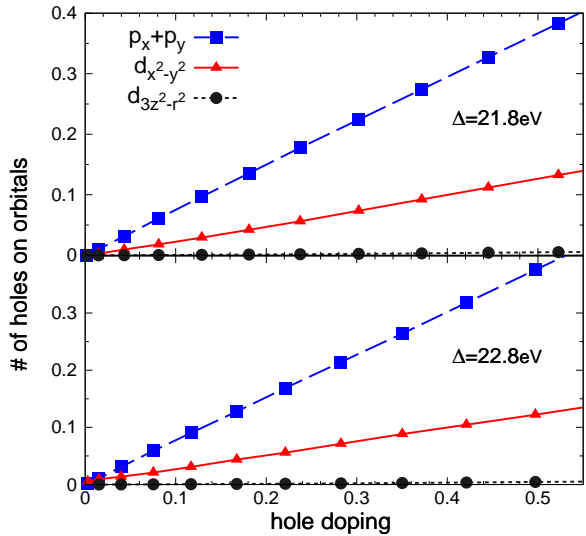}
    \caption{Number of holes on each orbital as a function of the total hole doping per unit cell, calculated for the six-band model in $T'$-phase. Upper panel: $\Delta=21.8$ eV ($<\Delta_{c2}$). Lower panel: $\Delta=22.8$ eV ($=\Delta_{c1}$). Note that the number of holes on $p_x$ and $p_y$ orbital of the in-plane oxygen sites are combined as $p_x+p_y$ (shown as blue squares with long dashed lines).}
    \label{numholesnopz}
\end{figure}

\begin{figure}[]
    \centering
    \includegraphics[width=8cm, angle=0]{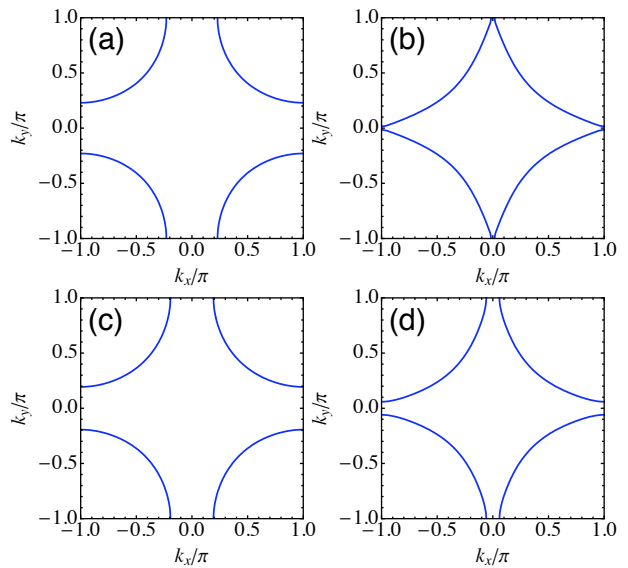}
    \caption{Fermi Surfaces of the six-band model in $T'$-phase. Panel (a): $\Delta=21.8$ eV ($<\Delta_{c2}$), hole doping $x=0.13$. Panel (b): $\Delta=21.8$ eV, $x=0.37$. Panel (c): $\Delta=22.8$ eV ($=\Delta_{c1}$), $x=0.12$. Panel (d): $\Delta=22.8$ eV, $x=0.35$. Parameters: $U=9$ eV, $T=0.1$ eV. 
    Panel (a): $\varepsilon_d=\varepsilon_{d_z}=-25.3$ eV, $\varepsilon_p=-3.5$ eV.
     Panel (b): $\varepsilon_d=\varepsilon_{d_z}=-24.5$ eV, $\varepsilon_p=-2.7$ eV.
      Panel (c): $\varepsilon_d=\varepsilon_{d_z}=-26.1$ eV, $\varepsilon_p=-3.3$ eV.
       Panel (d): $\varepsilon_d=\varepsilon_{d_z}=-25.3$ eV, $\varepsilon_p=-2.5$ eV.}
    \label{fermisurfacenopz}
\end{figure}

We have repeated the entire analysis for the $T'$ structure, finding very similar results but with even smaller occupancy of the $d_{3z^2-r^2}$ orbitals. This is understandable as the hybridization to the $d_{3z^2-r^2}$ orbital is much weaker once $p_z$ orbitals are removed. We also plot the Fermi Surface in Fig.~\ref{fermisurfacenopz}. As for the $T$-phase, the Fermi surface has only one sheet, and the $d_{3z^2-r^2}$ orbitals are not populated as separate bands. We note that the van Hove singularity happens at around hole doping value $x\approx0.37$ which is approximately a 0.1 shift in doping  compared to the $T$ phase.

\section{Conclusion}\label{sec:conclusion}

In this paper, we have employed the single-site DMFT method to study a six-band model, which includes copper $d_{x^2-y^2}$, $d_{3z^2-r^2}$, in-plane oxygen $p_{x,y}$, and (in $T$ phase structure) the apical oxygen $p_z$ orbitals. This model is more chemically realistic than the three-band or one-band models frequently considered.  We considered two structures: the $T$-phase, appropriate to La$_2$CuO$_4$, and the $T'$-phase, appropriate to the infinite-layer cuprates and to the electron-doped materials such as Nd$_2$CuO$_4$.  From the model point of view these structures different in whether or not  apical oxygen $p_z$ orbitals are incluced. We have mapped out the metal/charge-transfer-insulator phase diagram, finding that after the atomic-limit Hartree shift is  accounted for, the phase boundaries are systematically shifted to the metallic regime compared to that of the three-band model. Thus we conclude that incorporating the $d_{3z^2-r^2}$ orbital expands the insulating regime of the system. 

The spectral functions are calculated by analytic continuation. The $d_{x^2-y^2}$ and $p_{x,y}$ spectra are observed to be similar to that of the three-band model. In the $T$-phase the non-bonding $p_z$ band appears as a $\delta$-function and two side-bands corresponds to the bonding apical oxygen bands. Hybridization to these orbitals means that the  $d_{3z^2-r^2}$ has a double-peak structure. In contrast, in the $T'$-phase, the spectrum of $d_{3z^2-r^2}$ orbital has a single peak. In the insulating regime, we have found that the insulating gap is generically smaller in the six-band model than in the three-band model, for comparable correlation parameters.

We have calculated the $d$-$d$ exciton spectrum, finding a sharp exciton line which should be visible in experiments. In the insulating phase, the exciton states are inside the charge-transfer gap. In the metallic phase, the exciton states are at slightly higher binding energy, but although they overlap in energy with the tails of the Hubbard bands, the broadening is small.

Both in-plane and $c$-axis optical conductivy are calculated. We have found, in disagreement with previous publication,\cite{Weber10b} that inclusion of the additional $d_{3z^2-r^2}$ and apical oxygen bands does not fix the problem of the near-gap magnitude. The $c$-axis conductivity is much weaker and the absorption is very small in the frequency range 2 eV to 4 eV. Above 4 eV there is a  relatively noticeable absorption due to transition from the decoupled apical oxygen bands to the upper Hubbard band.

We have studied the distribution of doped holes onto different orbitals. We have shown that under no physically relevant doping values that the $d_{3z^2-r^2}$ orbitals (and $p_z$ orbitals in the $T$-phase) are populated as a separate band. The Fermi surface only contains one sheet.  This is in disagreement with previous slave boson studies.\cite{Feiner92}

\section*{Acknowledgments}
We thank M. Capone for helpful discussions, and N. Lin for cross-checking our results with exact diagonalization solvers.  XW is supported by the Condensed Matter Theory Center of University of Maryland, and HTD and AJM by NSF-DMR-1006282.
Part of this research was conducted at the Center for Nanophase Materials Sciences, which is sponsored at Oak Ridge National Laboratory by the Division of Scientific User Facilities, U.S. Department of Energy. The impurity solver is based on a code primarily developed by P. Werner\cite{Werner06b} and uses the ALPS library.\cite{ALPS}

\appendix
\renewcommand{\theequation}{A-\arabic{equation}}
\setcounter{equation}{0}
\section*{Appendix}

In this appendix we present the ``six-band+s'' model involving an additional Cu $4s$ orbital in each unit cell. We take the basis as $|\psi\rangle=\left({d_{
\parallel\boldsymbol{k}}},
{d_{z\boldsymbol{k}}}, {s_{\boldsymbol{k}}}, {p_{x\boldsymbol{k}}}, {p_{y\boldsymbol{k}}},
{p_{z\boldsymbol{k}}^{\rm above}}, {p_{z\boldsymbol{k}}^{\rm below}}\right)$. Then the
Hamiltonian is a $7\times7$ matrix, which may be seperated to Cu and
O parts as
\begin{equation}
{\bf H}_{\rm 6band+s}=\left(\begin{array}{cc}
{\bf H}_{\rm 6band+s}^{\rm Cu} & {\bf H}_{\rm 6band+s}^{\rm hyb} \vspace{0.1cm}\\
\left({\bf H}_{\rm 6band+s}^{\rm hyb}\right)^\dagger & {\bf
H}_{\rm 6band+s}^{\rm O}
\end{array}\right),\label{sixbplusswhole}
\end{equation}
where
\begin{equation}
{\bf H}_{\rm 6band+s}^{\rm Cu}=\left(\begin{array}{ccc}
\varepsilon_d & 0 & 0\\
0 & \varepsilon_{d_z} & 0\\
0 & 0 & \varepsilon_{s}
\end{array}\right),
\end{equation}
\begin{equation}
{\bf H}_{\rm 6band+s}^{\rm O}=\left(\begin{array}{cccc}
\varepsilon_p & 0 & 0 & 0\\
0 & \varepsilon_p & 0 & 0\\
0 & 0 & \varepsilon_{p_z}& 0\\
0 & 0 & 0 & \varepsilon_{p_z}
\end{array}\right),
\end{equation}
and the hybridization between Cu and O orbitals:
\begin{equation}
\begin{split}
&{\bf H}_{\rm 6band+s}^{\rm hyb}=\\
&\left(\begin{array}{cccc}
2it_{pd}\sin\frac{k_x}{2} & -2it_{pd}\sin\frac{k_y}{2} & 0 & 0\\
-2it_{pd_z}\sin\frac{k_x}{2} & -2it_{pd_z}\sin\frac{k_y}{2} & t_{p_zd_z} & -t_{p_zd_z}\\
2it_{ps}\sin\frac{k_x}{2} & 2it_{ps}\sin\frac{k_y}{2} & t_{p_zs} &
-t_{p_zs}
\end{array}\right).
\end{split}
\end{equation}

The downfolding\cite{Loewdin51} of Eq.~\eqref{sixbplusswhole} leads to the Hamiltonian considered in the main text. The effective oxygen-oxygen hopping amplitudes are 
\begin{align}
 t_{pp}=\frac{t_{ps}^2}{\varepsilon_s-\varepsilon_F}\\
 t_{pp_z}=\frac{t_{ps}\cdot t_{p_zs}}{\varepsilon_s-\varepsilon_F}\\
 t_{p_zp_z}=\frac{t_{p_zs}^2}{\varepsilon_s-\varepsilon_F}
\end{align}
Note that this implies that  $t_{pp_z}/t_{pp}=t_{p_zp_z}/t_{pp_z}$ which has been used in obtaining the value of $t_{p_zp_z}$ in the main text.

\end{document}